\documentclass[conference, 9pt]{IEEEtran}
\usepackage{comment}
\usepackage{tikz}
\usepackage{subcaption}
\usepackage[hidelinks]{hyperref}
\usepackage{listings}
\usepackage{amsfonts,amsmath}
\usepackage{import}

\usepackage{threeparttablex}
\usepackage[cjk]{kotex}
\usepackage{enumerate, float, pgf, circuitikz, multirow, longtable, pdfpages, hyperref, array,siunitx, diagbox}
\usepackage{calc}
\usepackage{pgfplots}
\usepackage{xtab,booktabs}
\usepackage{lipsum}
\usepackage{makecell}
\usepackage{graphicx}
\usepackage{subcaption}
\usepackage{threeparttable}
\usepackage{pifont}

\usepackage{colortbl}
\usepackage{tabularx,booktabs}

\usetikzlibrary{arrows,positioning,calc,fit,shapes,decorations.markings}
\usetikzlibrary{shapes.arrows}

\definecolor{CommentGreen}{rgb}{0,.6,0}
\definecolor{numbercolour}{gray}{0.5}
\definecolor{keywordc}{rgb}{.63,0,.42}
\definecolor{comment}{rgb}{0,.6,0}
\definecolor{keyword}{rgb}{.63,0,.42}
\definecolor{kw2}{rgb}{.50,.50,.15}
\definecolor{kw3}{rgb}{.42,.42,.63}
\definecolor{string}{rgb}{1,0,0}

\lstdefinestyle{vcode}{
	escapechar=@@,
	basicstyle=\ttfamily\footnotesize,
	language=Verilog,
	frame=tb,
	numbers=left,
	fontadjust=true,
	basewidth=0.5em,
	showstringspaces=false,
	keywordstyle=\bfseries\color{magenta},
	commentstyle=\color{CommentGreen},
	tabsize=4,
	morekeywords={always_ff,always_comb,always_latch,priority,unique,%
	interface,modport,enum,struct,foreach,logic,typedef,bit,packed,unpacked,%
	tagged,union,endinterface,\$dumpfile,\$dumpvars%
	}
}
\lstdefinestyle{scode}{
	basicstyle=\ttfamily\footnotesize,
	language=Lisp,
	frame=tb,
	numbers=left,
	fontadjust=true,
	basewidth=0.5em,
	showstringspaces=false,
	keywordstyle=\bfseries\color{magenta},
	commentstyle=\color{CommentGreen},
	tabsize=4,
	morekeywords={if,define_insn,match_operand,include,unspec_volatile,restrict},
}
\lstdefinestyle{ccode}{
	basicstyle=\ttfamily\footnotesize,
	language=C++,
    escapechar=\$,
	frame=tb,
	numbers=none,
	fontadjust=true,
	basewidth=0.5em,
	showstringspaces=false,
	keywordstyle=\bfseries\color{magenta},
	commentstyle=\color{CommentGreen},
	tabsize=4,
    prebreak=\raisebox{0ex}[0ex][0ex]{\ensuremath{\hookleftarrow}},
	morekeywords={mmap,mprotect,clone,futex,__asm__,uint32_t,override,constexpr,fork}
}

\DeclareFontFamily{U}{skulls}{}
\DeclareFontShape{U}{skulls}{m}{n}{ <-> skull }{}
\DeclareRobustCommand{\skull}{\ifmmode\mathskull\else\textskull\fi}
\newcommand\textskull{{\usefont{U}{skulls}{m}{n}\symbol{'101}}}
\makeatletter
\newcommand\mathskull{%
  \check@mathfonts
  \mathchoice
    {\mbox{\fontsize{\tf@size}{\z@}\textskull}}
    {\mbox{\fontsize{\tf@size}{\z@}\textskull}}
    {\mbox{\fontsize{\sf@size}{\z@}\textskull}}
    {\mbox{\fontsize{\ssf@size}{\z@}\textskull}}%
}
\makeatother
\newcommand{\projectname}{HeisenTrojan}

\newcolumntype{R}{>{\raggedleft\arraybackslash}X}
\newcolumntype{L}{>{\raggedright\arraybackslash}X}

\makeatletter
\newcommand{\linebreakand}{%
  \end{@IEEEauthorhalign}
  \hfill\mbox{}\par
  \mbox{}\hfill\begin{@IEEEauthorhalign}
}

\begin{document}

\IEEEoverridecommandlockouts
\IEEEpubid{\begin{minipage}[t]{\textwidth}\ \\[10pt]
        \centering\normalsize{979-8-3503-4099-0/23/\$31.00  \copyright 2023 IEEE}
\end{minipage}} 

\title{\projectname{}s: They Are Not There Until They Are Triggered}

\author{\IEEEauthorblockN{Akshita Reddy Mavurapu}
\IEEEauthorblockA{University of New Hampshire \\
akshitareddy.mavurapu@unh.edu}
\and
\IEEEauthorblockN{Haoqi Shan}
\IEEEauthorblockA{CertiK \\
haoqi.shan@certik.com}
\and
\IEEEauthorblockN{Xiaolong Guo}
\IEEEauthorblockA{University of Kansas \\
guoxiaolong@ksu.edu}
\and
\IEEEauthorblockN{Orlando Arias}
\IEEEauthorblockA{University of Massachusetts, Lowell \\
orlando\_arias@uml.edu}
\linebreakand
\IEEEauthorblockN{Dean Sullivan}
\IEEEauthorblockA{University of New Hampshire \\
dean.sullivan@unh.edu}
}

\maketitle

\begin{abstract}

The hardware security community has made significant advances in detecting Hardware Trojan vulnerabilities using software fuzzing-inspired automated analysis. However, the Electronic Design Automation (EDA) code base itself remains under-examined by the same techniques. Our experiments in fuzzing EDA tools demonstrate that, indeed, they are prone to software bugs.
As a consequence, this paper unveils \projectname{} attacks, a new hardware attack that does not generate harmful hardware, but rather, exploits software vulnerabilities in the EDA tools themselves. A key feature of \projectname{} attacks is that they are capable of deploying a malicious payload on the system hosting the EDA tools without triggering verification tools because \projectname{} attacks \emph{do not} rely on superfluous or malicious hardware that would otherwise be noticeable.
The aim of a \projectname{} attack is to execute arbitrary code on the system on which the vulnerable EDA tool is hosted, thereby establishing a permanent presence and providing a beachhead for intrusion into that system. Our analysis reveals 83\% of the EDA tools analyzed have exploitable bugs. In what follows, we demonstrate an end-to-end attack and provide analysis on the existing capabilities of fuzzers to find \projectname{} attacks in order to emphasize their practicality and the need to secure EDA tools against them. 

\end{abstract}

\section{Introduction}

Recently, significant effort in the hardware security community has been paid to the automated analysis of Hardware Trojan vulnerability detection~\cite{kande2022thehuzz, trippel2022fuzzing} by borrowing concepts from software fuzzing. Less attention, however, 
has been paid in applying those same concepts to the analysis of the EDA code base itself. EDA tools are complex and sophisticated pieces of software comprising millions of lines of code (MLoC) and heavily used in the community. Moreover, there is a clear correlation between the number of lines of code, third-party library reliance, and number of users for a tool on the one hand and the number of errors and reported vulnerabilities on the other \cite{woody2014predicting, grimes2015beware}. This insight is, in part, what led us to ask if we \emph{can we expect that the EDA tool itself is bug free.}

Our results indicate that, similar to other existing complex code bases, EDA tools contain buggy code. We evaluate a representative set of common EDA tools and found an exploitable bug in 83\% of tools analyzed, other instances of bugs that may be exploitable given a motivated attacker with enough time, and still other instances of bugs that, while not fully exploitable for a full system compromise, can be used as means of rendering the tool unusable (Denial of Service).

\begin{figure}[t]
    \centering
    \includegraphics[width=\linewidth]{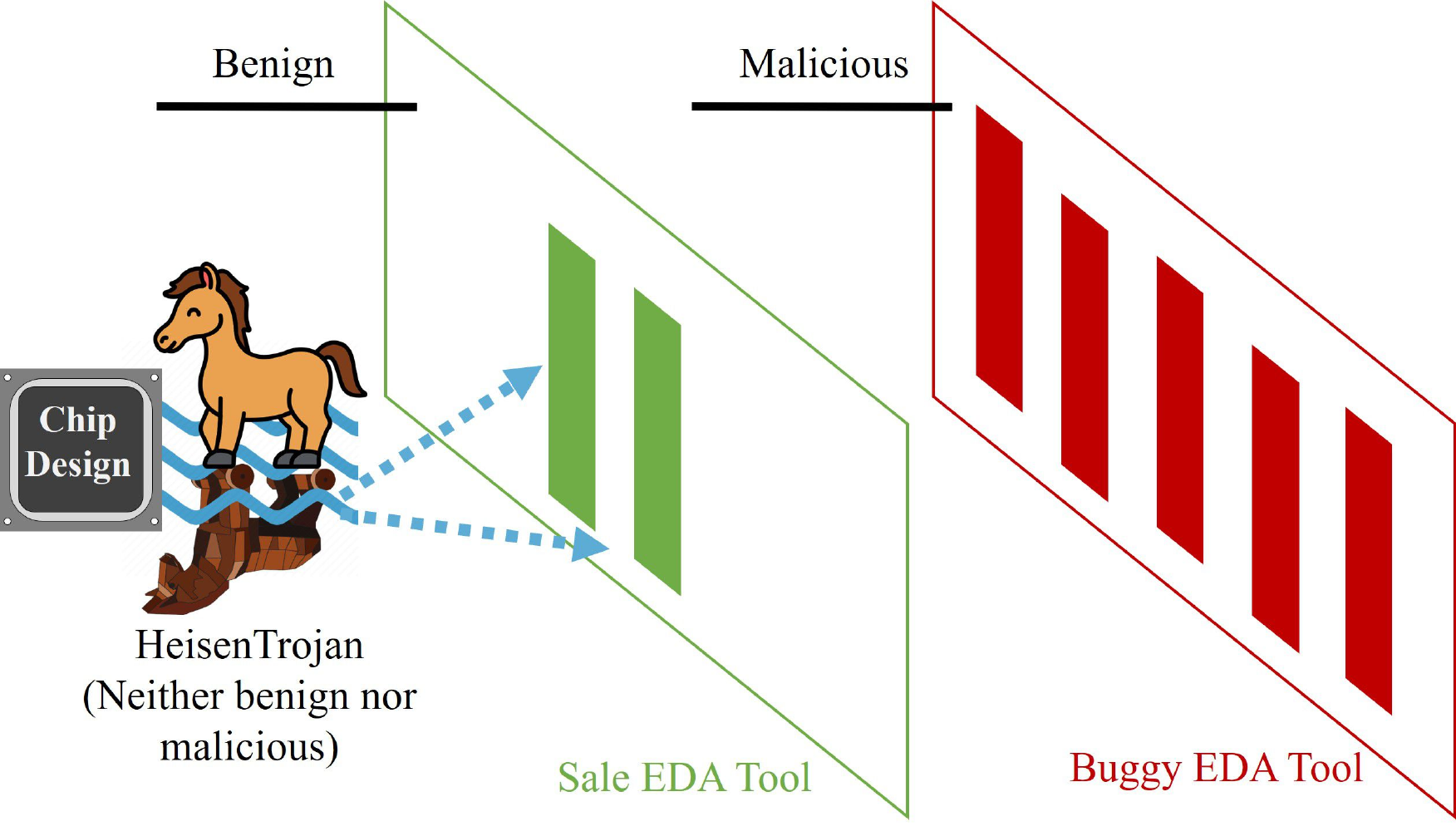}
    \caption{Borrowing from the famous double slit experiment, a \projectname{} projected onto an EDA tool exhibits benign properties if the EDA tool is safe, but is malicious if the EDA tool is vulnerable.}
    \label{fig:heisentrojan}
    \vspace{-0.6cm}
\end{figure}

In this paper we present a new class of hardware attacks called \textbf{\projectname{}s}. To begin, we consider these hardware attacks because the malicious input is embedded in hardware-related components such as a hardware descrption language (HDL), simulation test vectors, and waveforms. As such, we are using ``hardware" in the broad sense to include anything typically involved in the EDA toolchain and hardware manufacturing life cycle. While we say throughout the paper that, for instance, the HDL in a \projectname{} is malicious, it is not inherently so as it \emph{does not} generate any malicious or superfluous hardware.
Instead, as shown in Figure \ref{fig:heisentrojan}, a \projectname{} aims to exploit vulnerabilities only on buggy EDA tools to deploy a malicious payload in the system running the tool in order to compromise it. This duality property is the inspiration behind the name.

The goal of a \projectname{} attack is to gain arbitrary code execution on the machine hosting the vulnerable EDA tool. This, in itself, typically the first step for an attacker who wishes to obtain persistence as in a supply-chain attack \cite{cao2022fork}.
For ethical considerations, in this work we do not show the precise steps to achieve a full system compromise. We do, however, use this as justification to show how a \projectname{} can be a significant threat to organizations and companies by presenting how a full \projectname{}-based attack can be developed. Further, we consider \projectname{}s relevant for two additional reasons.

First, \projectname{} attacks represent a new attack vector to the hardware security community. We do not claim that triggering software vulnerabilities to gain arbitrary code execution is new, but rather, that targeting software vulnerabilities in EDA tools via maliciously crafted inputs is new. Concern in the hardware security community is typically focused on preventing and/or detecting maliciously crafted HDL that attempts to embed a traditional Hardware Trojan. Little attention, if any, has been paid to the EDA tools themselves\footnote{Recent work explored attacks stemming from malicious EDA tools but focused on their ability, in such a scenario, to surreptitiously embed hardware Trojans \cite{sunkavilli2021analysis}. More will be said regarding this, and other, related works in Section \ref{sec:related}.}. We show that this should be a concern. Further, it is an unexplored attack surface by virtue of its newness. To the best of our knowledge, this is the first paper to report an end-to-end attack targeting an EDA tool. 

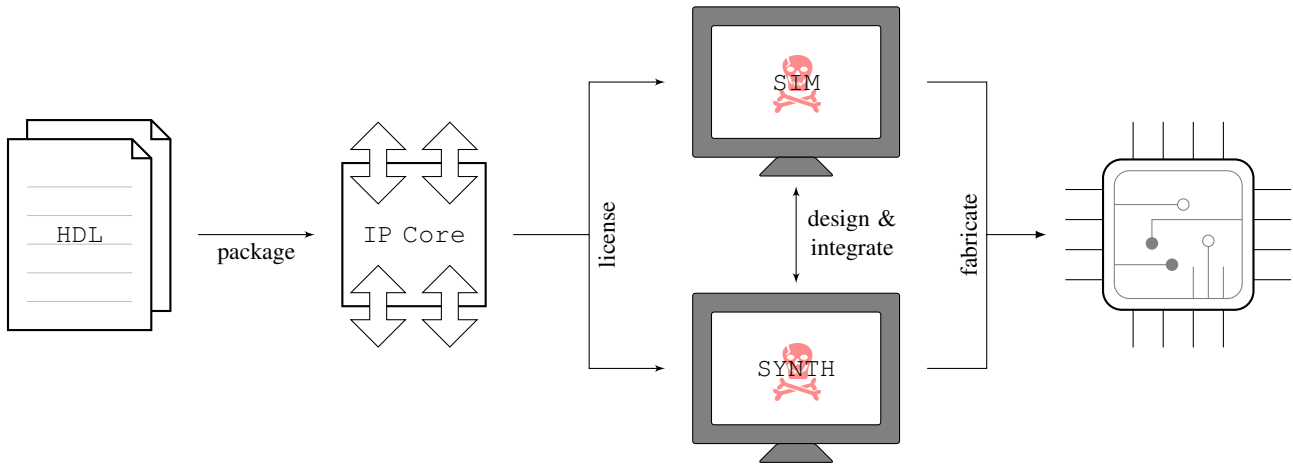
\begin{figure*}[h]
    \centering
    \begin{tikzpicture}[
        vsheet/.style = {
            font=\ttfamily\large,
            text width=0.75in,
            minimum height=1in,
            text centered,
        },
    ]

    \begin{scope}
        \begin{scope}[shift={(0.1in, 0.1in)}]
            \draw[thick, fill=white] (-0.375in, -0.5in) -- ++(0.75in, 0)
                -- ++(0, 0.9in) -- ++(-0.1in, 0.1in)
                -- ++(-0.65in, 0) -- cycle
                (0.375in, 0.4in) -| ++(-0.1in, 0.1in);
        \end{scope}
        \draw[thick, fill=white] (-0.375in, -0.5in) -- ++(0.75in, 0)
            -- ++(0, 0.9in) -- ++(-0.1in, 0.1in)
            -- ++(-0.65in, 0) -- cycle
            (0.375in, 0.4in) -| ++(-0.1in, 0.1in);
        \foreach \y in {1,...,5} {
            \draw[very thin, gray!50] (-0.275in, 0.15in*\y -0.5in) -- ++ (0.55in, 0);
        }
        \node[vsheet] at(0, 0) (verilog) {HDL};
    \end{scope}

    \begin{scope}[shift={(1.75in, 0in)}]
        \draw[thick] (-0.375in, -0.375in) rectangle ++(0.75in, 0.75in);
        \foreach \x in {0,1} {
            \node[draw, double arrow, fill=white, minimum height=0.05in, text width=0.25in, rotate=90]
                at(0.375in * \x -0.1875in, 0.375in) {};
            \node[draw, double arrow, fill=white, minimum height=0.05in, text width=0.25in, rotate=90]
                at(0.375in * \x -0.1875in, -0.375in) {};
        }
        \node[vsheet, minimum height=0.75in] (ip)
            at(0, 0) {IP Core};
    \end{scope}

    \begin{scope}[shift={(3.75in, 0.75in)}]
    	\draw[color=black, fill=gray, rounded corners=1]
    		(-1.375, 1.125) rectangle (1.375, -0.875);
    	\draw[color=black, fill=white, rounded corners=1]
    		(-1.125, 0.875) rectangle (1.125, -0.625);
    	\draw[color=black, fill=gray, rounded corners=1]
    		(-0.25, -0.875) -- (-0.5, -1.125) -- (0.5, -1.125) -- (0.25, -0.875) -- cycle;
        \node[font=\Huge, color=red!45] at(0, 0.125) {\skull};
        \node[font=\ttfamily\large, text width=2.75cm, minimum height=2.25cm, text centered] at(0, 0.125) (sim) {SIM};
    \end{scope}

    \begin{scope}[shift={(3.75in, -0.75in)}]
    	\draw[color=black, fill=gray, rounded corners=1]
    		(-1.375, 1.125) rectangle (1.375, -0.875);
    	\draw[color=black, fill=white, rounded corners=1]
    		(-1.125, 0.875) rectangle (1.125, -0.625);
    	\draw[color=black, fill=gray, rounded corners=1]
    		(-0.25, -0.875) -- (-0.5, -1.125) -- (0.5, -1.125) -- (0.25, -0.875) -- cycle;
        \node[font=\Huge, color=red!45] at(0, 0.1) {\skull};
        \node[font=\ttfamily\large,text width=2.75cm, minimum height=2.25cm, text centered] at(0, 0.125) (synth) {SYNTH};
    \end{scope}

    \begin{scope}[shift={(5.75in, 0)}]
        \draw[thick, rounded corners=0.25cm]
            (-1, -1) rectangle (1, 1);
        \draw[gray, rounded corners=0.2cm]
            (-0.85, -0.85) rectangle (0.85, 0.85);
        \foreach \x in {1,2,3,4} {
            \draw ({2*\x/5 - 4/5 - 2/10}, -1) -- ++(0, -0.5);
            \draw ({2*\x/5 - 4/5 - 2/10},  1) -- ++(0,  0.5);
            \draw ( 1, {2*\x/5 - 4/5 - 2/10}) -- ++(0.5,  0);
            \draw (-1, {2*\x/5 - 4/5 - 2/10}) -- ++(-0.5, 0);
        }
        \draw[thin, gray, -o] (0.4, -0.85) -- ++(0, 0.85);
        \draw[thin, gray] (0.6, -0.85) -- ++(0, 0.425);
        \draw[thin, gray] (0.2, -0.85) -- ++(0, 0.425);
        \draw[thin, gray, -*] (0.85, 0.2) -| ++(-1.2, -0.4);
        \draw[thin, gray, -*] (-0.85, -0.4) -- ++(0.85, 0);
        \draw[thin, gray, -o] (-0.85,  0.4) -- ++(1, 0);
        \coordinate (die) at (-1.5, 0);
    \end{scope}

    \draw[-latex'] ($ (verilog.east) + (0.2in, 0) $) --
        node[anchor=north] {package} ( $ (ip.west) - (0.1in, 0) $ );

    \draw[latex'-latex'] ( $(sim.south) - (0, 0.1in)$) --
        node[anchor=west, text width=4em, text centered] {design \& integrate}
            (synth);

    \draw[-latex'] let \p1=(ip.east) in let \p2=(sim.west) in let \p3=(synth.west) in
        (\x1 + 0.1in, \y1) -| ({(\x1 + \x2)/2}, \y2) -- (\x2 - 0.1in, \y2);
    \draw[-latex'] let \p1=(ip.east) in let \p2=(sim.west) in let \p3=(synth.west) in
        ({(\x1 + \x3)/2 + 0.75em}, \y1) node[rotate=90] {license}
        ({(\x1 + \x3)/2}, \y1) |- (\x3 - 0.1in, \y3);

    \draw[-latex] let \p1=(sim.east) in let \p2=(die) in let \p3=(synth.east) in
        (\x3 + 0.1in, \y3) -| ({(\x3 + \x2)/2}, \y2)
        ({(\x3 + \x2)/2 - 0.75em}, \y2) node[rotate=90] {fabricate}
        (\x1 + 0.1in, \y1) -| ({(\x1 + \x2)/2}, \y2) -- (\x2 - 0.1in, \y2);
    
\end{tikzpicture}
    \caption{The \projectname{} makes use of standard HDL workflows in the industry. Malicious HDL is packed as part of an IP core. This HDL does \emph{not} aim to generate malicious hardware. Instead, it triggers and exploits bugs in an HDL synthesis or simulation tool in order to comprome the system where the tool runs. If fabricated, ICs including \projectname{} affected IP cores do not exhibit malicious behavior.}
    \label{fig:overview}
    \vspace{-0.5cm}
\end{figure*}

Second, \projectname{}s can be practically exploited in a variety of relevant scenarios. We will present the details of an end-to-end attack in Section \ref{sec:exploitable_bug}. As shown in Figure \ref{fig:overview}, at a high-level, we reason that an attacker can leverage the HDL design, development, and integration process. The first relies on the complexity of HDL designs. For instance, we found we can hide the offending input in a large design, with MLoC, that will unlikely be analyzed line-by-line. Furthermore, by their nature, \projectname{}s do not generate malicious or superfluous hardware and thus go undetected by verification tools. The second scenario makes use of the need to guard intellectual property (IP) by releasing it to third-party vendors \cite{speith2022not}. In such a scenario, it is trivial to embed a maliciously crafted input which can exploit vulnerabilities in EDA tools. The third scenario exploits shared computing environments. FPGA resources are currently scant and expensive. It is not uncommon to turn to cloud providers who offer/lease relatively cheap FPGA resources to users in a hosted environment as an alternative. This allows the attack to arbitrarily upload\footnote{Within reason and as prescribed by the cloud vendors rules \cite{aws2023fpga}.} malicious input to a cloud-hosted EDA tool.  

\subsection{Motivation}

We aim to address several motivating points in introducing \projectname{} attacks that we hope will further aid its understanding:

\medskip
\noindent\textbf{How did we discover \projectname{} attacks?}
We were inspired by recent efforts that have demonstrated significant results from the automated analysis of HDL %
using software fuzzing techniques. However, it is not our aim to improve those results, but rather, to query the EDA tools themselves by treating them purely as pieces of software. For instance, an EDA tool accepts HDL as input for synthesis and a combination of HDL and test vector input during simulation. In each case, we are interested in finding vulnerabilities in the way the EDA tool handles those inputs when guided by a mutational fuzzer.

\medskip
\noindent\textbf{Why do we target synthesis, simulation, and scripting?} Generally, EDA tools take an HDL design and perform synthesis and place-and-route to generate a bitstream that can then be used to configure an FPGA, or alternatively target a technology library to generate an ASIC design. Simulation is performed prior to place-and-route and uses a testbench as a wrapper to handle the flow of input and output to/from the design under test. These two cases allow us to mutate input either in the form of synthesizable HDL or as a test vector input. We also target traditional EDA tool scripting languages because they often act as a front-end to automate the design flow from synthesis to place-and-route, or other types of
analysis.

\subsection{Contributions}

To summarize, our contributions include:
\begin{itemize}
    \item We present \projectname~attacks, a unique hardware attack that is benign if the EDA tool is bug-free but malicious if it is not.
    \item We highlight the practicality of \projectname~attacks by presenting an end-to-end exploit for an EDA tool.
    \item We present analysis of the effectiveness of fuzzing EDA tools in the discovery of \projectname~attacks.
    
\end{itemize}

In what follows, we will first introduce relevant background in Section \ref{sec:background} before discussing related works in Section~\ref{sec:related}. We then present an end-to-end \projectname{} in Section \ref{sec:attack}. Analysis of the use of fuzzing to find \projectname{}s is presented in Section \ref{sec:fuzzing} prior to a discussion of future work and conclusions in Section \ref{sec:conclusion}.

\section{Background}\label{sec:background}
    \subsection{EDA Tools}
We evaluate 6 EDA tools in total including two synthesis tools (\texttt{yosys} \cite{wolf2016yosys}, \texttt{abc} \cite{brayton2010abc}), three simulation tools (\texttt{iverilog} \cite{williams2002icarus}, \texttt{verilator}~\cite{snyder2004verilator}, and \texttt{gtkwave} \cite{bybell2010gtkwave}), and a formal analysis tool (\texttt{z3} \cite{de2008z3}). It is assumed that the readers are familiar with these tools, so we forego a detailed explanation of their functionality and internals. Instead, we provide a high-level overview of them in Table \ref{table:eda_tools} and refer the reader to their source for further information.

\begin{table}[h]
\centering
\caption{Open-source EDA tools analyzed in this work.}
\label{table:eda_tools}
\begin{tabularx}{\columnwidth}{ %
                                lX
                                }
\toprule    
\textbf{EDA Tool} & \textbf{Description} \\
\midrule
iverilog & A verilog HDL compiler for the IEEE-1364 standard \\
gtkwave & A GTK+~based wave viewer \\
yosys & Open source synthesis suite for Verilog\\
abc & A synthesis and verification suite \\
z3 & An SMT solver developed by Microsoft \\
verilator & A verilog HDL simulator \\
\bottomrule
\end{tabularx}
\end{table}

We focus our evaluation on open source EDA tools because they are easily obtained at no cost and are supportive of bug reporting. We have reported our results to the tool maintainers. However, they are still in the patching process so, in this paper, we limit our discussion to number of bugs found and type. We remove all details of the EDA tool, insofar as possible, in the end-to-end exploit to abide with responsible disclosure procedures. 

\subsection{Fuzzing}
Fuzzing is one of the most successfully employed techniques for software bug discovery \cite{liang2018fuzzing} and currently an active area of research in both the hardware and software communities. Fuzzers are usually categorized as black-box, white-box, or grey-box depending on the amount of information they have of the underlying program. A black-box fuzzer knows nothing about the internal structure of the program, whereas a white-box fuzzer knows everything about the program's internal structure. A grey-box fuzzer sits in-between the two in that it has limited knowledge of the internals of the program behavior via coverage-guided feedback provided by some form of program instrumentation. 

In this paper, we rely heavily on coverage-guided fuzzing. Briefly, the fundamental steps taken in coverage guided fuzzing include:

\begin{enumerate}
    \item \textbf{Instrumentation:} The program being fuzzed is augmented with code to record its control-flow during execution.
    \item \textbf{Seed pool Generation:} The fuzzer generates candidate inputs for the program, which can either be user supplied or randomly generated. The correct generation of an initial seed pool is an active area of research~\cite{hazimeh2020magma}. 
    \item \textbf{Input mutation:} After each round, selected inputs that exhibit unique code coverage are mutated (e.g. addition, deletion of bytes or flipping, rotating bits among several others).
    \item \textbf{Coverage-guided feedback:} Code coverage information is recorded as the seed inputs are executed. This information is binned, ordered, and then selected (e.g. best/new code coverage) to generate the new input corpus for the next fuzzing round.
\end{enumerate}

Using open-source tools allows us to use gray-box fuzzers with little effort. Closed-source EDA tools require the use of black-box fuzzers which are slow and unreliable. Since our objective is to demonstrate the dangers of \projectname{}s we believe this is a fair compromise.

\section{Related Works}\label{sec:related}
    We are unaware of any existing effort to fuzz EDA tools and craft \projectname{} attacks. Project F4PGA \cite{murray2020symbiflow,kashifchipshop} aims to reverse engineer bitstreams for Lattice and Xilinx FPGAs using what amounts to fuzzing, but has a completely different goal than our work. However, we are not claiming complete ingenuity. Analysis of software using a fuzzer is common-place and likely performed in-house by EDA tool vendors, or at the very least during the development process for reporting and fixing bugs prior to release. With that said, we are only aware of one bug report \cite{vivadobug}, but there are probably many that go unreported publicly.

Recent research \cite{herklotz2020finding}  investigates logic synthesis tools for the correctness of their output via equivalence checking. The research is constructed around a tool, Verismith, that generates semantically correct and deterministic Verilog allowing for comparison between the generated design and its synthesized netlist. If they differ, then a bug is registered. A similar analysis is performed for high-level synthesis tools \cite{herklotz2021empirical}. Our work, however, is not interested in the correctness of output for a given logic synthesis tool, but rather, whether a given EDA tool contains bugs that can triggered via specially crafted input to gain control of the system. We show that many such bugs can indeed be found for a variety of commonly used tools (synthesis, simulation, verification) and present an end-to-end exploit to demonstrate their impact. 

An orthogonal line of research was proposed in \cite{sunkavilli2021analysis, sunkavilli2023fpga} that investigates the attack surface of malicious EDA tools. The authors demonstrate the potential for attacks by, for instance, making minor modifications to the intermediate files generated by the EDA toolchain. This research highlights and provides insight into the vulnerabilities that can be exploited by attackers if the EDA toolchain is compromised. An interesting approach may be to combine \projectname{}~attacks with this line of research to first gain control of the EDA tool in order to then embed a traditional Hardware Trojan.

\section{Building a \projectname{}}\label{sec:attack}
    In what follows, we first introduce our adversarial model. We then discuss the steps taken to build a full end-to-end \projectname{} attack chain on an EDA tool. Due to page limitations, we are limited to discussing a single attack against a given EDA tool. However, we found 12 exploitable bugs in total, see Table~\ref{table:bugs}, in which we are able to craft end-to-end exploits across the tools outlined.

\subsection{Adversarial Model and Assumptions}
\projectname{} introduces a new class of hardware attacks where an adversary can ship an IP core written in any given HDL, a simulation test vector, or script for tool management that contains a specially crafted payload. However, unlike traditional Hardware Trojans the bundle in question \emph{does not} generate any malicious hardware. Instead, the adversary's goal is to \emph{compromise the computing system of the licensee}. This may be done with the purpose of spying, sabotaging operations, or stealing data. The \projectname{} shipped as part of the IP core exploits a vulnerability in the tooling utilized by the victim to establish a permanent presence in their computing system. When processed by a non-vulnerable tool, the \projectname{} is not triggered nor does it generate extraneous or superfluous hardware, thereby remaining hidden from detection.

\subsection{Finding an Exploitable Bug}
\label{sec:exploitable_bug}
We examined the results of our fuzzing campaign on open-source EDA tools until we found a crash that was the result of an error that could be readily exploited. Suitability for exploitation was determined by examining the crash and performing a two factor test:
\begin{itemize}
    \item whether the behavior causing the crash is controllable in ways that do not trigger a denial of service; and
    \item whether the behavior causing the crash allows for the corruption of memory areas containing code pointers.
\end{itemize}
A suitable vulnerability was found in tool \emph{REDACTED} \footnote{We have followed responsible disclosure procedures for all vulnerabilities found and reported them to the respective maintainers. The tool name is omitted to comply with procedures.}. The vulnerability allows us to perform a write anywhere in the program's stack, which is conducive to a type of attack under the umbrella of code-reuse attacks called \emph{return-oriented programming} (ROP) \cite{shacham2007geometry,roemer2012return}.

\subsection{Preparing, Deploying, and Synthesizing the HDL}
For purposes of deployment the HDL can be wrapped in IEEE 1735-2023 \cite{10123338} but this is not strictly necessary. The goal is \emph{not} to hide a traditional Hardware Trojan which generates malicious logic, but to deploy HDL which generates legitimate hardware. %
The \projectname{} payload aims to exploit bugs on the EDA toolchain being used to compromise the victim's infrastructure. As such, for our purposes, the HDL can be sent in plain-text, as long as the portion of the HDL which triggers the bug in the HDL tool is innocuous.%

\begin{lstlisting}[style=ccode,caption={Simplified view of the vulnerability.},label={lst:vulnerable_code}]
char buffer[1024], * p = buffer;
va_list ap;
/* ... */
p += vsnprintf(p, buffer+sizeof(buffer)-p, fmt, ap); /* $\ding{182}$ */
p += snprintf(p, buffer+sizeof(buffer)-p, "\n");     /* $\ding{183}$ */
\end{lstlisting}

We make use of a write anywhere vulnerability we found in the implementation of the \lstinline[style=ccode]{f_REDACTED_r()} function of one of the synthesis tools we tested. By crafting a specific payload and exploiting this vulnerability we are able to overwrite a return address in the program's stack, thereby allowing us to perform a control-flow attack. Through repeated usages of this vulnerability, we craft a payload which results in a chain of gadgets that achieve our desired result. In our case, we wish to print an innocent message in the terminal to signify a successful attack.

When synthesized, the HDL in question makes use of one of the vulnerabilities presented in Section \ref{sec:fuzzing}, Table \ref{table:bugs}. A simplified version of the bug is shown in Listing \ref{lst:vulnerable_code}. The code in question attempts to perform safe string concatenation through the use of the \lstinline[style=ccode]{snprintf()} family of functions. The next available location in the buffer is found by advacing the pointer \lstinline[style=ccode]{p} by the return of the \lstinline[style=ccode]{vsnprintf()} function in the line labeled \ding{182}. However, it is imperative to notice that these functions \emph{do not} return the value of characters added to the buffer, but the number of characters that \emph{would have been added} were there enough room in the buffer. As such, on the call to \lstinline[style=ccode]{vsnprintf()} in \ding{183} the pointer may point to an address outside the bounds of the buffer. This gives us a \emph{spatial} memory error which can be exploited to perform a write to a location of memory.

The spatial memory error in question allows us to overwrite any byte on memory to have a value of \lstinline[style=ccode]{0x00}. Because the buffer in question is allocated in the stack of a function automatic storage variables and any code pointers stored in the stack, such as return addresses, are prime targets. Since the vulnerability in question allows us to freely move a pointer to any address in the stack area, we can safely bypass commonly deployed defenses such as stack protection \cite{wagle2003stackguard} while constructing the desired ROP-chain. This allows us to gain arbitrary code execution by chaining gadgets (e.g. small instruction sequences) together from the existing code base.  

\subsection{A Word about Full System Compromise}
We show how we can achieve arbitrary code execution by exploiting bugs in the synthesis environment. A system compromise relies on a payload which can cause permanent changes to the OS introducing malware. Achieving this goal further requires escalation of privileges through a kernel vulnerability. The latter are common \cite{CVE-2023-35788,CVE-2023-3812,CVE-2023-4004,CVE-2023-4147} and do not require our payload to directly exploit them. Our payload can simply launch another application (such as through the \lstinline[style=ccode]{execve()} family of functions) which can more readily exploit such vulnerabilities.

\section{On Fuzzing EDA Tools}\label{sec:fuzzing}
    \begin{table}
\centering
\caption{Number, type, and exploitability of bugs per EDA tool.}
\label{table:bugs}
\begin{threeparttable}[h]
\begin{tabularx}{\columnwidth}{ %
                                lcXr
                                }
\toprule    
\textbf{EDA Tool} &
\textbf{Bugs\tnote{$\dagger$}} &
\textbf{Types} &
\textbf{Vulnerable\tnote{$\ddagger$}} \\
\midrule
\multirow{1}{*}{Z3} & \multirow{1}{*}{1}     & null pointer dereference & \ding{53} \\ 
\midrule[0.5pt]
\multirow{2}{*}{GTKWave} & \multirow{2}{*}{7} 
    & heap overflow 
    & \ding{52} \\
    & & null pointer dereference 
    & \ding{53} \\
\midrule[0.5pt]
\multirow{2}{*}{verilator} & \multirow{2}{*}{3}
    & stack overflow
    & \ding{52} \\
    & & null pointer dereference
    & \ding{53} \\
\midrule[0.5pt]
\multirow{2}{*}{iverilog} & \multirow{2}{*}{11}
    & stack overflow
    & \ding{52}\\
    & & null pointer dereference
    & \ding{53} \\
\midrule[0.5pt]
\multirow{1}{*}{ABC} & \multirow{1}{*}{9} 
    & null pointer dereference 
    & \ding{53} \\
\midrule[0.5pt]
\multirow{3}{*}{Yosys} & \multirow{3}{*}{6} 
    & heap overflow 
    & \ding{52} \\
    & & stack overflow 
    & \ding{52} \\
    & & null pointer dereference 
    & \ding{53} \\
\midrule
\textbf{Total} & \textbf{37} & & \textbf{12} \\
\bottomrule
\end{tabularx}
\begin{tablenotes}
    \item[$\dagger$] Number listed corresponds to unique bugs.
	\item[$\ddagger$] Indicates if a bug creates a vulnerability which can be used to perform an attack other than a Denial of Service.
\end{tablenotes}
\end{threeparttable}
\vspace{-0.5cm}
\end{table}

\begin{figure*}[ht]
    \centering
    
    \begin{subfigure}[b]{.3\linewidth}
        \includegraphics[width=0.99\textwidth]{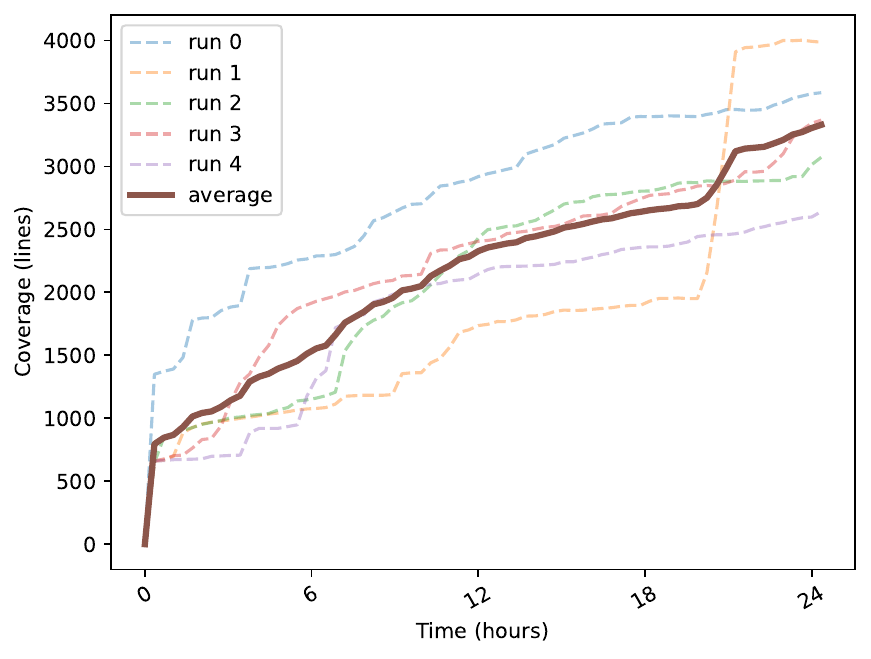}
        \caption{ABC}\label{fig:1a}
    \end{subfigure}
    \begin{subfigure}[b]{.3\linewidth}
        \includegraphics[width=0.99\textwidth]{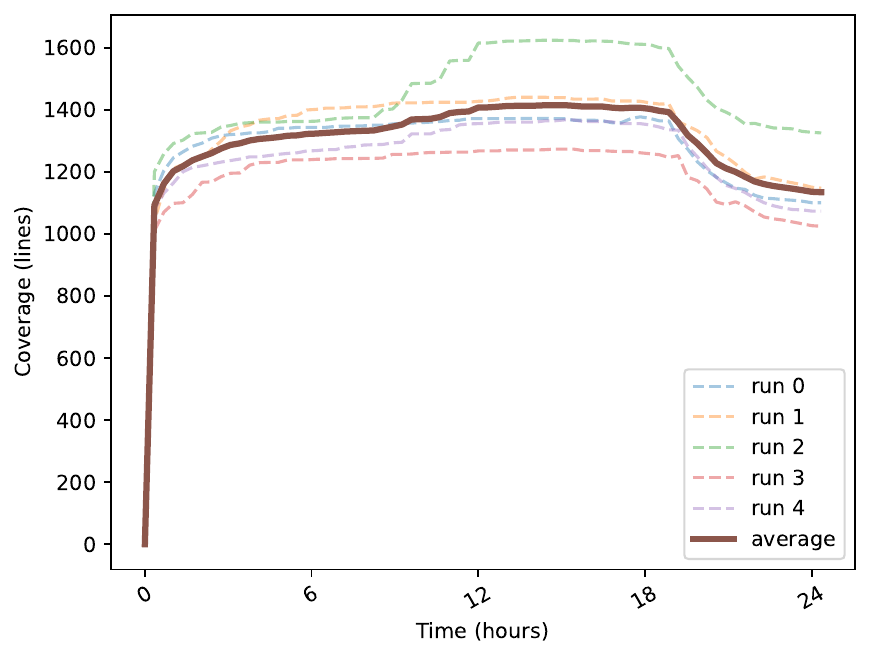}
        \caption{GTKWave}\label{fig:1b}
    \end{subfigure}
    \begin{subfigure}[b]{.3\linewidth}
        \includegraphics[width=0.99\textwidth]{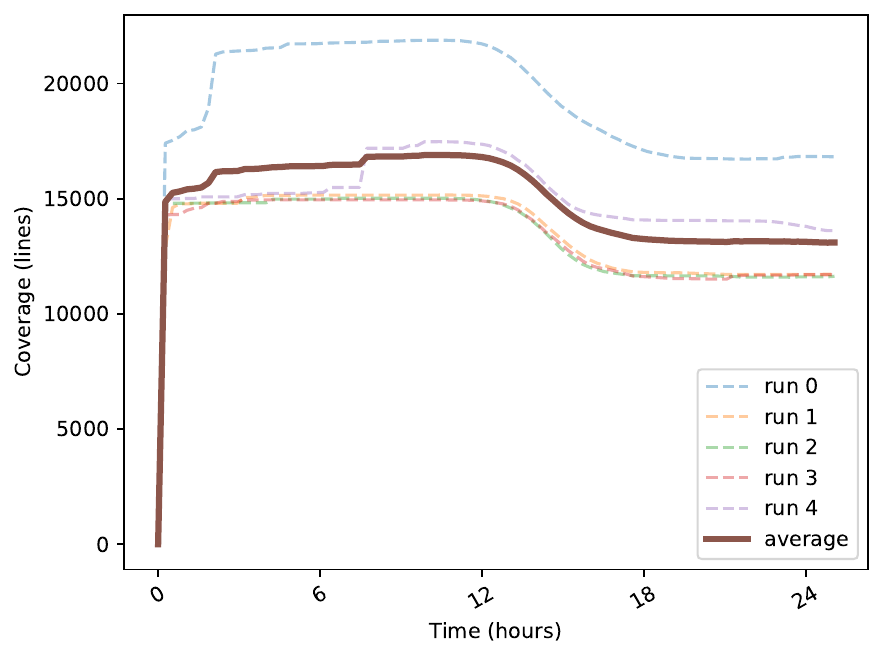}
        \caption{z3}\label{fig:1c}
    \end{subfigure}
    \begin{subfigure}[b]{.3\linewidth}
        \includegraphics[width=0.99\textwidth]{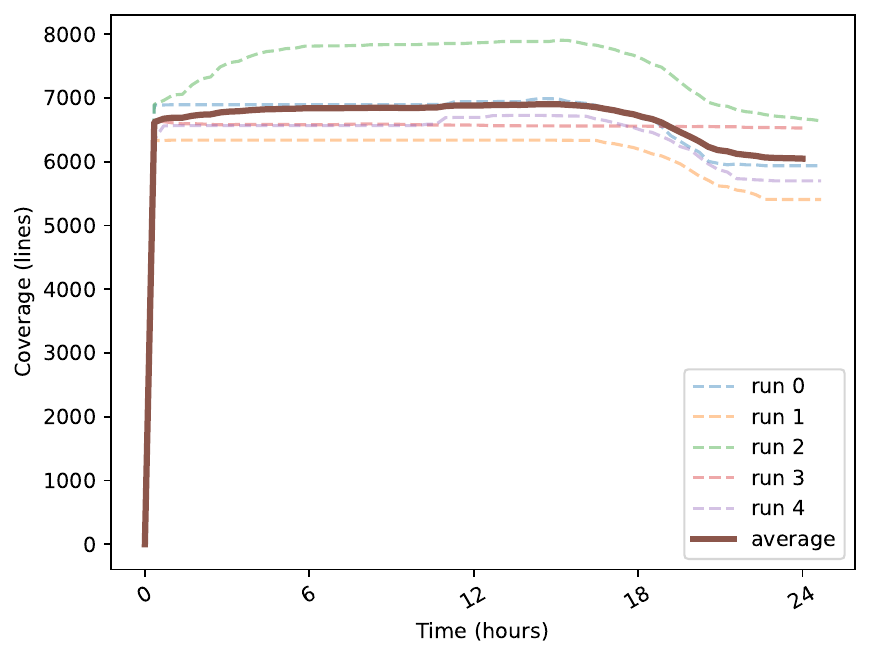}
        \caption{yosys}\label{fig:1d}
    \end{subfigure}
    \begin{subfigure}[b]{.3\linewidth}
        \includegraphics[width=0.99\textwidth]{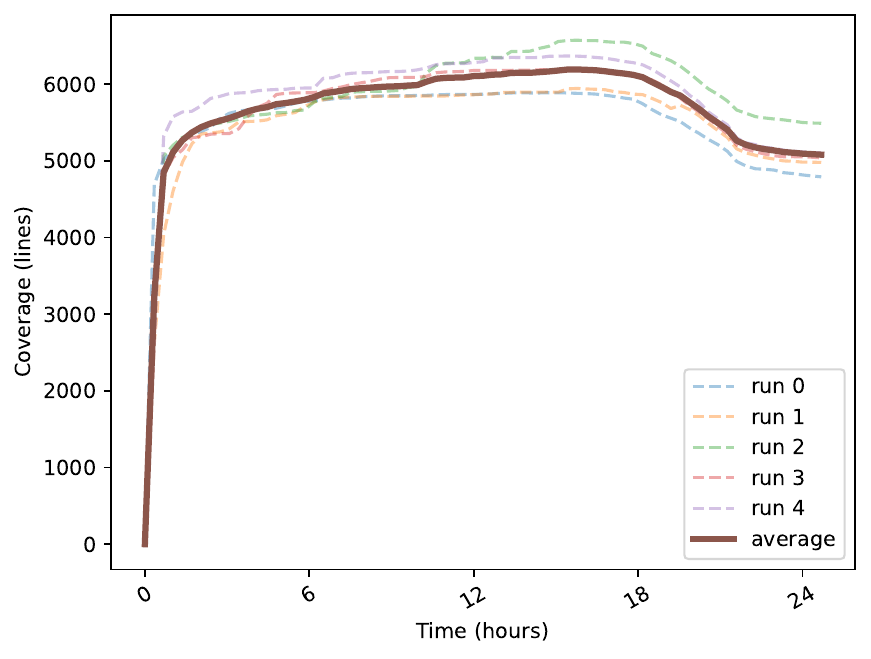}
        \caption{iverilog}\label{fig:1e}
    \end{subfigure}
    \begin{subfigure}[b]{.3\linewidth}
        \includegraphics[width=0.99\textwidth]{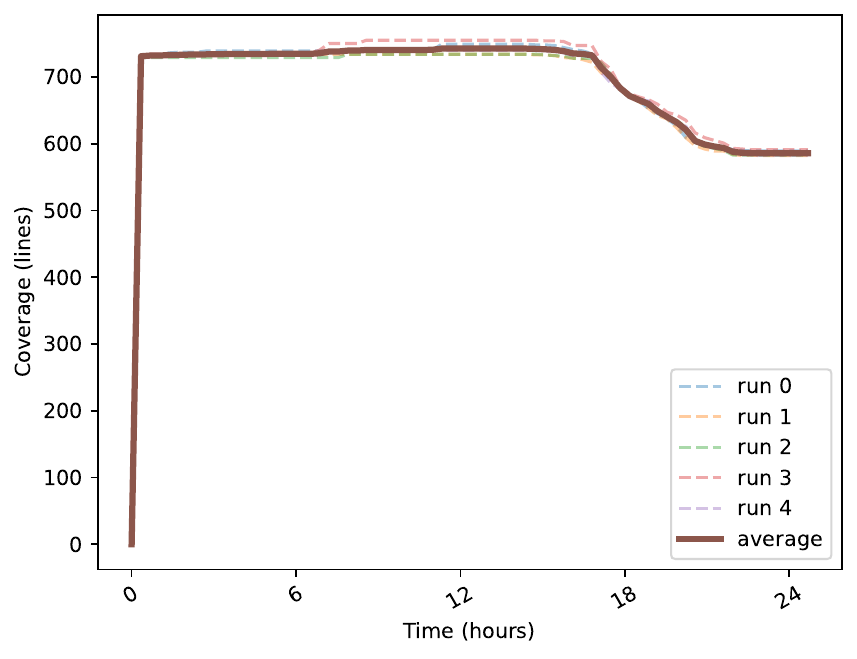}
        \caption{verilator}\label{fig:1f}
    \end{subfigure}
    \caption{Coverage plots for evaluated EDA tools.}
    \label{fig:coverage}
    \vspace{-0.5cm}
\end{figure*}

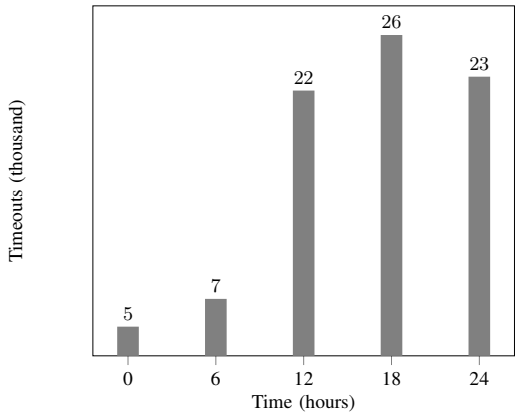
\begin{figure}[htb]
    \centering
        \resizebox{0.8\columnwidth}{!}{
            \begin{tikzpicture}  
  
\begin{axis}  
[xtick pos=left, ytick=\empty,
    ybar,   
    ylabel={Timeouts (thousand)},  
    xlabel={Time (hours)},  
    symbolic x coords={0, 6, 12, 18, 24},  
    xtick=data,  
     nodes near coords,   
    nodes near coords align={vertical},  
    ]  
\addplot[fill=gray,draw=none] coordinates {(0,5) (6,7) (12,22) (18,26) (24,23) };  
  
\end{axis}  
\end{tikzpicture} 
         }
    \caption{Number of timeouts per fuzzing interval for z3}
    \label{fig:timeouts}
    \vspace{-0.5cm}
\end{figure}

In what follows, we present a discussion on how we found bugs on each tool we examined as well as a description of the process and tooling used. We summarize the bugs found in Table \ref{table:bugs} as well as present an analysis of coverage achieved.

\subsection{Experimental Tools}
We used \texttt{honggfuzz} \cite{swiecki2017honggfuzz} to fuzz each EDA tool due to its comparative performance in recent literature \cite{li2021unifuzz,metzman2021fuzzbench}. Honggfuzz is a multi-threaded, grey-box coverage-guided fuzzer. It takes input corpora, checks them for new coverage, and then feeds the files to a simple in-memory corpus directory. It utilizes randomly chosen inputs from this memory and mutates them to start fuzzing. 

We used \texttt{llvm-cov} \cite{llvm2023cov} to generate coverage information for each EDA tool. The tools are instrumented to emit profile and coverage information by compiling them with \texttt{clang} using the \texttt{-fprofile-instr-generate} and \texttt{-fcoverage-mapping} flags. Coverage information is generated by running the instrumented EDA tool normally. A raw profile file is created that is then converted using the \texttt{llvm-profdata merge} tool. Finally, a \texttt{JSON}-formatted file is exported to collect the profiled edge coverage during the measured interval.

We used a combination of \texttt{AFLTriage} \cite{afltriage} and \texttt{afl-tmin} \cite{fioraldi2020afl++} to determine the uniqueness of the bugs found. 
\texttt{AFLTriage} is first used to minimize the number of crashing inputs by performing crash deduplication using a heuristic that selects between file and line number and the address of the first interesting stack frame. This approach is imperfect and may lead to missing unique crashes (false negatives) but avoids labelling aliasing crashes as unique (false positives). \texttt{afl-tmin} is then used to minimize a crashing input to its bare minimum (in terms of bytes). This allows us to more easily determine how to control the bug by modifying the minimized input.  

\subsection{Experiment Setup}
We instrumented the source code for each EDA tool, used a trivial input seed (e.g. an appropriate ``hello, world'' for the tool), and applied flags where necessary to ensure proper functionality of the tool with the fuzzer (e.g. disabling \texttt{gtkwave's} GUI) while fuzzing. 
Each EDA tools was evaluated 5 times for a 24 hour period on a 24 core Intel i7-12850HX with 32 GiB of memory running Ubuntu 22.04.2 LTS for a total of 20,000 compute hours per recommendations \cite{metzman2021fuzzbench}.

\subsection{Unique Crashes and Crash Types}

Table \ref{table:bugs} shows the number of unique crashes, crash types, and exploitability of the crash for each EDA tool. In total, we found 37 unique crashes. Limited manual analysis was used to determine that 12 of these bugs were exploitable from userland. This does not mean that the remaining bugs are not exploitable, but rather they were not immediately exploitable given our analysis. A motivated attacker under the correct conditions could exploit the bugs.

We categorized the bugs broadly as either null-pointer dereferences, heap, or stack overflows. With respect to the latter two classes, we made no distinction between a out-of-bounds access or an overflow for  clarity. We found, however, that the heap overflows were typically write-anywhere vulnerabilities that crashed while accessing invalid memory. Of these, we found that 62\% were controllable - we could read/write to an arbitrary location in mapped memory. 

\subsection{Evaluation of Coverage}
Coverage plots for each EDA tool are provided in Figure~\ref{fig:coverage}. It is apparent from these plots that, in the majority of cases, the fuzzing campaign shows diminishing returns as it progresses in time. This can be explained, in part, as a side-effect of the method of logging coverage information. After each logging interval, the saved corpora in the default directory used by the fuzzer are removed and saved in an accumulated directory of corpora from prior intervals. This was done to avoid analyzing prior corpora for coverage information and, thereby, save time. However, we also reason that the fall-off in coverage over time is due to the highly structured input requirements for EDA tools. The initial seed, while trivially constructed for our evaluation, represents a valid starting point for the fuzzer. This can be seen by the immediate increase in line coverage. However, as time elapses, that input becomes less structured under mutation by the fuzzer. We observed that those programs which show diminishing returns experience a concomitant increase in number of timeouts. This effectively prevents the fuzzer from making forward progress during each interval of coverage evaluation. 

Only one tool, \texttt{abc}, shows steady growth. Others show almost no growth after initialization. The \texttt{z3} SMT solver performs worst in that it shows diminishing returns after only 12 hours of fuzzing. This makes sense considering its strict input requirements. We plot the number of timeouts over time for \texttt{z3} as a topical explanation for these results in Figure~\ref{fig:timeouts}. Notice that the number of timeout increases at 12 hours. We observed a similar trend for the other EDA tools too. 

\begin{table}[h]
\centering
\caption{Percent line coverage achieved while fuzzing compared to total line coverage.}
\label{table:percent}
\begin{tabularx}{\columnwidth}{Xcc
                                }
\toprule    
\textbf{EDA Tool} & \textbf{12h (\%)} & \textbf{24h (\%)}   \\ 
\midrule
iverilog & 21 & 17 \\
GTKWave & 15 & 13 \\
Yosys & 11 & 9 \\
ABC & 15 & 23 \\
Z3 & 13 & 9 \\
verilator & 8 & 5 \\
\bottomrule
\end{tabularx}
\vspace{-0.1cm}
\end{table}

We also show the achieved line coverage compared to total line coverage as a percentage in Table \ref{table:percent}. 
In most cases, our results indicate that the fuzzer was only capable of shallow analysis. Improving the seed corpus would perhaps improve these results per discussions in related work \cite{hazimeh2020magma}. However, a more obvious, yet more complicated, solution would be to develop a custom interface capable of handling highly-structured input for the fuzzer.

\section{Future Work and Conclusions}\label{sec:conclusion}
    There is a clear trend between a tool's code base and complexity of its user ecosystem, and the number of reported memory vulnerabilities for that tool \cite{woody2014predicting, grimes2015beware}. \projectname{} attacks exploit this trend to launch end-to-end attacks against common EDA tools. Therefore, in order to defend against them, an EDA tool vendor could eliminate all memory vulnerabilities in its code base. Research, however, suggests that its unlikely memory corruption can be completely eliminated from runtime efficient languages without paying a severe performance loss \cite{szekeres2013sok} or switching entirely to a memory safe language \cite{fulton2021benefits}. The latter is unlikely, though gaining traction, and the former untenable. Another mitigation strategy against \projectname{} attacks includes code-reuse mitigations. However, these too are often imperfect or negatively impact performance \cite{larsen2018continuing}.

Another promising strategy would be to incorporate fuzzing into the EDA tooling ecosystem. As Section \ref{sec:fuzzing} demonstrated, current fuzzing infrastructure is ill-suited to handle the highly structured input requirements for the common EDA tools. An interesting area of research, partially explored in \cite{herklotz2020finding}, includes the creation of domain-specific front-ends capable of outputting correctly formatted data. This would provide a sane mutational block with which the fuzzer can work. This would not only resolve the diminishing returns observed during our experimentation, but also allow deeper exploration of the code base with respect to coverage. 

To conclude, in this paper we introduced a new class of Hardware Trojan which we call \projectname{}s. Unlike traditional hardware attacks, \projectname{}s do not generate spurious or malicious hardware but aim to exploit vulnerabilities in EDA tools to infect and establish a presence on computer systems. We caution that \projectname{}s can be used as vectors to enable spying, sabotaging, or stealing data of a potential rival or organization by attacking their computing infrastructure. We further showcase how, with some effort, \projectname{}s can be developed to target a wide array of EDA tools. Although in this work we focus on open-source tools, commercial closed-source tools are likely not bug-free and vulnerable to the same types of attack. We hope that this work raises awareness of this attack vector and jolts the industry into better securing their tools.

\vspace{-0.1in}

\bibliographystyle{IEEEtran}%
\bibliography{edafuzz}

\end{document}